\documentclass{appolb}
\usepackage{epsfig}

\begin{document}
\title{Universality of QCD traveling-waves with running coupling beyond leading logarithmic accuracy%
\thanks{Presented at the School on QCD, Low-x Physics and Diffraction,
Copanello, Calabria, Italy, July 2007.}%
}
\author{Guillaume Beuf\footnote{email: guillaume.beuf@cea.fr}
\address{Institut de Physique Th\'eorique,\\
CEA, IPhT, F-91191 Gif-sur-Yvette, France\\
CNRS, URA 2306}
}
\maketitle
\begin{abstract}
We discuss the solutions of QCD evolution equations with saturation in the high energy limit. We present a general argument showing that, in the running coupling case,
the Next-to-Leading-Logarithmic (NLL) and higher order terms are irrelevant for the \emph{universal} asymptotic features of the solutions. 
\end{abstract}

\section{Introduction}
The BFKL equation \cite{Lipatov:1976zz} is known to provide an incomplete description of the hadronic or nuclear collisions in the the high energy limit of QCD. That evolution equation relies indeed on the assumption of a dilute partonic content of incoming particles, but it inevitably leads to denser and denser ones. It also violates the Froissart bound and therefore unitarity. At high density, and hence at very high energy, coherent collective effects modify the BFKL equation by reducing the emission of additional soft gluons. That mechanism of gluon saturation tames the BFKL growth of the cross sections, and is expected to restore unitarity. Several evolution equations with saturation have been derived, like the B-JIMWLK \cite{Balitsky:1995ub,JIMWLK} or the BK \cite{Balitsky:1995ub,Kovchegov} equations, implementing nonlinear effects on top of the BFKL equation. However, the result discussed here is independent of the precise saturation mechanism at work.

The features of gluon saturation are well known at the leading logarithmic accuracy (LL)  with fixed QCD coupling. The saturation scale $Q_s(Y)$, which is the typical momentum scale for the onset of nonlinear effects, is related to the color correlation length in the transverse plane. Hence, that scale plays the role of an infrared cut-off for the gluons radiation. By analogy with the Fisher and Kolmogorov-Petrovsky-Piskounov equation, it has been understood \cite{Munier:2003vc} that gluon saturation also back-reacts onto the dilute linear regime. By linearity, the solutions of the fixed coupling BFKL equation can be decomposed as sum of wave solutions, interpreting formally the rapidity $Y$ as \emph{time} and $L=\log{k_T^2/\Lambda^2}$ as \emph{space position} for the wave ($k_T$ being the gluon's transverse momentum). If a saturation mechanism is at work, one of these waves solutions, the \emph{critical} one, is selected dynamically in a $L$ interval which is growing with $Y$. Therefore, the solution loses memory of its initial condition during the evolution, even in the linear domain. The asymptotic shape of the solution and the asymptotic evolution of the saturation scale are then \emph{universal}.
That selection mechanism provides a natural explanation for the \emph{geometric scaling} \cite{Stasto:2000er} observed in the HERA data, as the critical wave solution has precisely that scaling property.

One should also consider the effect of running QCD coupling and of higher logarithmic orders, in order to do reliable studies. A first step in that direction consists in replacing by hand the fixed coupling in the LL equation by a running coupling, taken for example at a scale equal to the parent gluon transverse momentum $k_T$. In that case, the mechanism of selection of the critical wave solution seems to hold. However, the running coupling leads to a quite different universal asymptotic behavior of the solution \cite{Munier:2003vc,Mueller:2002zm} (see \cite{Triantafyllopoulos:2008yn} for a review). The evolution is slower, and in particular the saturation scale behaves as $\log Q_s^2(Y) \propto \sqrt{Y}$, instead of $Y$ in the fixed coupling case.

On the other hand, one can introduce NLL contributions to the kernel, but
keeping the coupling fixed, so that the linear part of the evolution equation writes formally
\begin{equation}
 \partial_Y N(L,Y)= \bar{\alpha} \left[ \chi_{LL}(-\partial_L) + \bar{\alpha}\ \chi_{NLL}(-\partial_L)+ {\cal O}(\bar{\alpha}^2) \right] N(L,Y) \, ,\label{FCNLLBFKL}
\end{equation}
where $\bar{\alpha}=N_c\ \alpha_s / \pi$, and $N$ is the Fourier conjugate of the dipole target amplitude. In that case, the asymptotic behavior of the solution is similar to the one in the fixed coupling LL case, except that the value of the critical parameters are changed, and depends on the NLL kernel eigenvalues \cite{Enberg:2006aq}.

\section{Saturation with running coupling and a NLL kernel}

Let us now discuss the most complete case, with both running coupling and higher orders. 
The result presented here was first found in Ref.\cite{Peschanski:2006bm} and Ref.\cite{Beuf:2007cw}.
It is possible to resum some of the NLL and higher order contributions by letting the coupling run, for example at the scale $k_T$, \ie taking $\bar{\alpha}(k_T^2)=1/bL$. Then, Eq.(\ref{FCNLLBFKL}) rewrites
\begin{equation}
 \partial_Y N(L,Y)=\frac{1}{b L} \left[ \chi_{LL}(-\partial_L) + \frac{1}{b L} \chi^{(1)}(-\partial_L)+ {\cal O}\left(\left(\frac{1}{b L}\right)^2\right)\right] N(L,Y)\, .\label{RCNLLBFKL}
\end{equation}
Using the $\omega$-expansion method of Ref.\cite{Ciafaloni:1998iv}, one replaces Eq.(\ref{RCNLLBFKL}) by the effective equation
\begin{equation}
 \partial_Y N(L,Y)= \frac{1}{b L}\ \chi_{eff} (-\partial_L, \partial_Y) N(L,Y)\, ,\label{RCNLLEFFBFKL}
\end{equation}
with a kernel whose eigenvalues are
\begin{equation}
\chi_{eff} (\gamma,\omega) = \chi_{LL}(\gamma) \left[1 + \frac{\omega \ \chi^{(1)}(\gamma)}{\chi_{LL}^2 (\gamma)} + {\cal O}\left(\omega^2\right) \right] \, .\label{EFFKER}
\end{equation}
In the brackets in the previous expression, the terms of $n$-th order $\omega^n$ are supposed to have poles of order $n$ at most, at $\gamma=0$ and $\gamma=1$. Hence, that $\omega$-expansion is safe only when $\omega\ll\gamma,$ $1-\gamma$. Assuming that condition to be fulfilled, if one tries to solve Eq.(\ref{RCNLLEFFBFKL}) in the large $Y$ limit using Laplace transform, one gets the position of the saddle-point $\omega_s\sim Y^{-1/2}$.
Thus, for large enough $Y$ (in practice several rapidity units), the $\omega$-expansion will be safe, and we will have $\log N \propto \sqrt{Y}$ and $\log Q_s^2(Y) \propto \sqrt{Y}$.
Expanding the kernel of Eq.(\ref{RCNLLEFFBFKL}) around $\omega=0$ and changing variable $Y$ into $\sqrt{Y}$, one gets
\begin{eqnarray}
 \frac{ b L}{2 \sqrt{Y}}  \partial_{\sqrt{Y}} N(L,Y)&=& \chi_{eff} (-\partial_L, 0) \ N(L,Y)\nonumber\\
 &+&\left[\frac{1}{2 \sqrt{Y}} \dot{\chi}_{eff} (-\partial_L, 0) \partial_{\sqrt{Y}} + {\cal O}\left(\partial_{\sqrt{Y}}^2\right)\right] N(L,Y)\ ,\label{RCNLLBFKLEXP}
\end{eqnarray}
where $\dot{\chi}_{eff}$ stands for the derivative with respect to $\omega$.
Thanks to Eq.(\ref{EFFKER}), the first line in Eq.(\ref{RCNLLBFKLEXP}) is the LL BFKL equation with running coupling, and the second line contains the higher order terms. The relevant regime determining the universal asymptotic properties of the solution corresponds to $L\propto\sqrt{Y}\gg 1$, and more precisely to $\tau \sim Y^{1/6}\gg 1$, with $\tau=\log (k_T^2/Q_s^2(Y))$ being the geometric scaling variable. Solving Eq.(\ref{RCNLLBFKLEXP}) perturbatively at large $Y$, one finds that the three leading orders, corresponding to ${\cal O}(Y^0)$, ${\cal O}(Y^{-1/6})$ and ${\cal O}(Y^{-1/3})$,  features only terms from the first line of Eq.(\ref{RCNLLBFKLEXP}). The leading order give the dispersion relation between $\gamma$ and $\omega$. In the case of the critical wave solution, which will be the relevant asymptotic solution in the presence of gluon saturation, the second order selects $\gamma=\gamma_c$, solution of $\chi_{LL}(\gamma_c)=\gamma_c\ \chi_{LL}'(\gamma_c)$. It determines the leading term of $\log N(L,Y)$ and of $\log Q_s^2(Y)$. Their next-to-leading terms are given by the ${\cal O}(Y^{-1/3})$ order terms in Eq.(\ref{RCNLLBFKLEXP}). In $N(L,Y)$, it corresponds to the Airy diffusion factor. Up to that order, the NLL kernel has not been relevant.

Therefore, we have shown that in the running coupling case, the universal asymptotic solution in the linear regime derived in Refs.\cite{Munier:2003vc,Mueller:2002zm} is valid not only for LL evolution equations with saturation but also for NLL ones (or even all orders ones). NLL terms in the kernel can modify the solution only at subleading orders not calculated in Refs.\cite{Mueller:2002zm,Munier:2003vc}. For example, the third term in the large $Y$ expansion of $\log Q_s^2(Y)$, which will be of order ${\cal O}(Y^{-1/6})$, will be NLL dependant.

Saturation and running coupling are thus essential ingredients to understand the high energy limit of the partonic content of hadrons, whereas other NLL or higher order contribution are not. One should however notice that the behavior of the solutions of saturation equations is more subtle in the running coupling case, and not fully understood \cite{Beuf:2008mb}, by contrast to the fixed coupling case.

\end{document}